\documentclass[final,5p,times,number,twocolumn]{elsarticle}
\usepackage{float}
\usepackage[caption=false,justification=justified]{subfig}  
\usepackage{graphicx}
\usepackage[dvipsnames]{xcolor}
\usepackage[english]{babel}
\usepackage{bm}
\usepackage{amsmath}
\usepackage{amssymb}
\usepackage{amsthm}
\usepackage{mathrsfs}
\usepackage[font=scriptsize]{caption}
\usepackage{ragged2e}
\usepackage[font=small,labelfont=bf,justification=raggedright]{caption}
\usepackage[utf8]{inputenc}
\usepackage[tikz]{bclogo}
\usepackage[framemethod=tikz]{mdframed}

\newsavebox{\measurebox}
\usepackage[colorlinks=false]{hyperref}

\begin{document}
	
\begin{frontmatter}
		
\title{Effective loop cosmology is a mere symplectic modification of standard cosmology}
\author[IPN]{Abraham Espinoza-García
}\ead{aespinoza@ipn.mx}
\affiliation[IPN]{organization={Unidad Profesional Interdisciplinaria de Ingeniería Campus Guanajuato del Instituto Politécnico Nacional},
addressline={Av. Mineral de Valenciana No. 200, Col. Fraccionamiento Industrial Puerto Interior}, 
city={Silao de la Victoria},
postcode={36275}, 
state={Guanajuato},
country={Mexico}}
%
		
\begin{abstract}
It is shown that several prescriptions for the effective continuum limit of the flat Friedmann-Lemaitre-Robertson-Walker loop quantum cosmology  can be understood as \emph{the exact classical limit} of the Wheeler-DeWitt quantization of certain dynamical systems, which are themeselves  symplectic modifications of the standard Friedmann-Lemaitre-Robertson-Walker model. It is also demonstrated that a similar situation holds for two highly relevant anisotropic models. The required symplectic modifications are explicitly constructed. It is argued that this state of affairs makes the case for the associated Wheeler-DeWitt framework to be considered as an alternate quantum cosmology paradigm.
\end{abstract}
		
		
\end{frontmatter}

\section{Introduction}
Several proposals for a quantum theory of the gravitational field point to a breakdown of the continuum limit of spacetime. Indeed, a simple physical argument related to black holes shows that this might very well be the case \cite{sabine}. Two prominent candidates which predict such behavior are loop quantum gravity (LQG) and string/M theory.

Loop quantum cosmology (LQC) \cite{lqc-review, lqc-review2, chapter-lqc} is by now a well established framework attempting to describe gravitational systems with finite degrees of freedom in which quantum effects are expected to be relevant (i.e. the very early universe and black holes \cite{KS-ashtekar}). Succinctly speaking, LQC is an upgrade of the good old Wheeler-DeWitt quantum cosmology, in which the ideas and methods of full LQG \cite{lqg-review} (see also the chapters \cite{chapter-lqg, chapter2-lqg}) are meant to be taken into account at the minisuperspace approximation (extensions to the midisuperspace regime have also been put forward \cite{lqc-review}). One of the main successes that the LQC community claims to have achieved is the resolution of the cosmological and typical black hole singularities---due to the quantum geometry underlying the LQC paradigm. 

An alternate approach to quantum cosmology is the so-called noncommutative cosmology, in which ideas and methods of noncommutative gravity/field theory \cite{nekrasov, szabo} are implemented at the minisupespace approximation, resulting in a framework which can be directly associated to noncommutative quantum mechanics\cite{ncqm-review}. This second route is  naturally related to string/M theory \cite{nekrasov, szabo}. 

In order to fix notation and have some minimal background on which to elaborate, we proceed now to give a more proper (yet brief) treatment of these two quantum proposals. 
\subsection{The usual holonomization prescription}

We first state some facts pertaining the effective scheme of LQC (somewhat thorough accounts can be found in the standard reviews, e.g. \cite{lqc-review, lqc-review2}). As already mentioned, LQC is a proposed quantization of symmetry-reduced general relativity (GR) which mimics some of the chief ideas and methods of LQG. Briefly, the standard Hamiltonian formulation of GR is rewritten (with the help of a canonical transformation) in terms of the so called $SU(2)$ Ashtekar-Barbero connection, rewriting GR in a similar way as the gauge theories of particle physics. However, it is found that canonical quantization requires the holonomies of the Ashtekar-Barbero connection (and not the connection itself) to be considered as the true configuration field variable. Therefore, in quantum cosmology,  holonomies of the symmetric sector of the connection superspace are to be  regarded as the configuration variable. The resulting quantized framework is known as loop quantum cosmology. It is important to note that LQC is not obtained as a result of symmetry-reducing full LQG. An important consequence is that in full LQG on has noncommutativity among the conjugate momenta fields (electric fluxes), whereas in LQC such noncommutativity is not featured at all. Information is therefore lost when implementing the minisuperspace approximation underlying the LQC paradigm.

The corrections related to the loop quantization can be in the overall divided into holonomy-corrections and inverse triad-corrections, the former being somewhat more relevant. As a consequence of such corrections, the cosmological singularity is replaced by a quantum bounce. The so-called continuum (effective) limit of LQC is achieved by focusing on the holonomy-related quantum corrections. In the isotropic case, the classical geometro-dynamical variables which are selected by (effective) LQC are $b=\gamma\dot{a}/(aN)$ and $v=a^3$ (where $\gamma$ is the so called Barbaro-Immirzi parameter, with $a$ the scale factor and $N$ the lapse function). Considering additionally a free homogeneous field $\phi$, the corresponding gravity-matter phase space is hence defined by the Poisson structure (written in variables $z=(b,\phi,v,p_\phi)$) 
\begin{equation}
\mathsf{P}_z=4\pi G\gamma\, \partial_{b}\wedge\partial_{v}+\partial_{\phi}\wedge\partial_{p_{\phi}}.\label{PB}
\end{equation}
Whereas the Hamiltonian is written in these variables $z$ as ($N=1$)
\begin{equation}
\mathcal{H}_z=-\frac{3}{8\pi G\gamma^{2}}b^2 v+\frac{p^{2}_{\phi}}{2v}.\label{ham1}
\end{equation}
It has been shown (via analytical \cite{reviews-lqc} and numerical \cite{review-numlqc} investigations) that (in the isotropic case) the quantum corrections associated to LQC are very well approximated by performing the \textit{replacement} 
\begin{equation}
b\to\frac{\sin(\lambda b)}{\lambda},
\end{equation}
within the classical Hamiltonian, where $\lambda^{2}=4\sqrt{3}\pi\gamma\ell^{2}_{p}$ is the lowest eigenvalue (corresponding to quantum states compatible with the assumed spatially homogeneous and isotropic geometry) of the area operator in the full loop quantum gravity---and where $\ell_p$ is the Planck length. The resulting effective Hamiltonian $\mathcal{H}_{\mathrm{eff}}$ is given by (in variables $z$)
\begin{equation}
{\mathcal{H}_{\mathrm{eff}}}_z=-\frac{3}{8\pi G\gamma^{2}\lambda^{2}}\sin^{2}(\lambda b)v+\frac{p^{2}_{\phi}}{2v}.
\label{ham2}
\end{equation}
Thus, the effective scheme of LQC supports meaningful quantum corrections within a purely classical setup. This situation has  allowed to carry exploratory investigations of quantum effects on a wide range of cosmological backgrounds in this rather simplified theoretical framework.   

The equations of motion associated to the ``holonomized'' Hamiltonian \eqref{ham2} with the standard Poisson structure 
\begin{equation}
\mathsf{P}_z=4\pi G\gamma\, \partial_{b}\wedge\partial_{v}+\partial_{\phi}\wedge\partial_{p_\phi},\quad z=(b,v,\phi,p_\phi),
\label{PB2}
\end{equation}
are
\begin{align}
&\dot{b}=4\pi G\gamma\frac{\partial\mathcal{H}_{\mathrm{eff}}}{\partial v}=-\frac{3}{\gamma\lambda^{2}}\sin^{2}(\lambda b),\label{eom-beta}\\
&\dot{V}=-4\pi G\gamma\frac{\partial\mathcal{H}_{\mathrm{eff}}}{\partial b}=\frac{3}{\gamma\lambda}V\sin(\lambda b)\cos(\lambda b),\label{eom-vol}\\
&\dot{\phi}=\frac{\partial\mathcal{H}_{\mathrm{eff}}}{\partial p_{\phi}}=\frac{p_{\phi}}{v},\label{eom-phi}\\
&\dot{p}_{\phi}=-\frac{\partial\mathcal{H}_{\mathrm{eff}}}{\partial\phi}=0.\label{eom-p-phi}
\end{align}
Since $v=a^{3}$, $\frac{\dot{v}}{3v}=\frac{\dot{a}}{a}=:H$, then, taking into account the effective Hamiltonian constraint, $\mathcal{H}_{\mathrm{eff}}=0$, and the equation of motion for $v$, we have,
\begin{equation}
H^{2} :=\left(\frac{\dot{v}}{3v}\right)^{2}=\frac{8\pi G}{3}\rho\left(1-\frac{\rho}{\rho_{\mathrm{c}}}\right),
\label{mod-friedmann}
\end{equation}
where 
\begin{equation}
\rho=\frac{\dot\phi^2}{2}=\frac{p^{2}_{\phi}}{2v^{2}}=\frac{3}{8\pi G\gamma^{2}\lambda^{2}}\sin^{2}(\lambda b),
\end{equation} 
and  
\begin{equation}
\rho_{{c}}=\frac{3}{8\pi G\gamma^{2}\lambda^{2}}\simeq0.41\rho_{\mathrm p},
\end{equation}
is the maximum value that $\rho$ can take in view of the effective Hamiltonian constraint (with $\rho_{\mathrm p}$ the Planck density). Equation \eqref{mod-friedmann} is the so called \textit{modified Friedmann equation}.

In full LQC it is shown that $b$ takes values on the interval $\left(0,\pi/\lambda\right)$ \cite{lqc-review} (alternatively, on the interval $(-\pi/2\lambda,\pi/2\lambda)$), therefore, the volume function reaches a stationary point at $b=\frac{\pi}{2\lambda}$ (alternatively, at $b=0$). From the equation of motion (\ref{eom-beta}) it is established that $b$ is a decreasing function of time ($\dot b\leq0$), which in turn implies that such stationary point is only reached once (and that the maximum of the energy density function is also attained only once). From this monotonic behavior of $b$ it also follows that such stationary point corresponds to a minimum (since $\rho$ is always bounded, the stationary point associated to a vanishing volume does not feature in the solutions of the equations of motion). Thus, in the effective scheme of LQC, a minimum of the volume function is reached precisely once, and this minimum volume corresponds to the maximum of the energy density function. This state of affairs is summarized by the statement that singularity resolution is achieved through a (minimum-volume) quantum bounce. As already stated, this singularity resolution is one of the landmark features of LQC, and in this effective scheme it is generic: all solutions undergo a minimum-volume bounce exactly once. One could argue that singularity resolution is achieved due to the semiclassical assumptions underlying this effective scheme, but it has been shown that the quantum bounce is likewise ``robust'' in the full LQC (of the flat FLRW model with a free scalar field) \cite{lqc-review2}. 

Note that the modified Friedmann equation (\ref{mod-friedmann}) incorporates in a rather simple way the occurrence of the bounce (at $\rho=\rho_{\mathrm{c}}$). In the limit $\lambda\rightarrow0$ (no area gap) we recover the ordinary Friedmann equation ${H}^{2}=\frac{8\pi G}{3}\rho$.

By making use of the Hamiltonian constraint, the equation for $b$ can be decoupled and solved by direct integration. Upon substitution of the solution $b=b(t)$ in \eqref{eom-vol} we find the corresponding solution $v=v(t)$. Similarly, using solution $v=v(t)$ in \eqref{eom-phi} we get the solution $\phi=\phi(t)$. The solutions are
\begin{align}
&b(t)=\frac{1}{\lambda}\mathrm{arccot}\left(\frac{3t}{\gamma\lambda}\right),\label{sol-beta-lqc}\\
&v(t)=C_1\sqrt{\gamma^{2}\lambda^{2}+9t^{2}},\label{sol-v-lqc}\\
&\phi(t)=C_2+\frac{p_{\phi}}{3C_1}\mathrm{log}\left(3t+\sqrt{\gamma^2\lambda^2+9t^2}\right),\label{sol-fi-lqc}\\
&p_\phi(t)=p_\phi.
\end{align}

\subsection{The classical limit of $\theta-$deformed quantum mechanics}\label{ncqm}
The usual $\theta-$deformed quantum mechanics is defined by the algebra \cite{ncqm-review},
\begin{equation}
[\hat{q}^a,\hat{q}^b]=i\theta^{ab}\hat{1},\quad [\hat{q}^a,\hat{p}_b]=i\hbar\delta^{a}_b\hat{1},\label{nc-q-com-rel}
\end{equation}
where $[\theta^{ab}]$ is an antisymmetric constant real matrix (the remaining brackets being null). The most important consequence of these more general commutation relations being that new dispersion relations among configuration observables arise.

One could interpret the above algebra as arising from an attempt at a standard quantization of a mechanical system $(\tilde{\Gamma},\mathcal{H})$, defined by the symplectic manifold $\tilde\Gamma$ ($\mathrm{dim}\left(\tilde\Gamma\right)=2n$) and a certain Hamiltonian function $\mathcal H$, for which the trivialization $(U,z)$ of $\tilde\Gamma$ defined by coordinates $z=(q^a,p_b)$ is \emph{not} symplectic (i.e., $(U,z)$ is a chart which does not verify the Darboux theorem)\footnote{Note that there are several conventions in use, we take $\omega=\mathrm{d}p_{a}\wedge\mathrm{d}q^{a}$ as the standard canonical form in coordinates $(q^1,\dots,q^n,p_1,\dots,p_n)$, and we take the Poisson structure as the unique inverse $\mathsf{P}$ of $\omega$. Our convention agrees, in particular, with \cite{novikov}.}. Specifically, the Poisson structure $\tilde{\mathsf{P}}$ corresponding to $(\tilde\Gamma,\mathcal{H})$ is to take the form (in coordinates $z$)
\begin{equation}
\tilde{\mathsf{P}}_z=\theta^{ab}\partial_{q^a}\wedge\partial_{q^b}+\partial_{q^a}\wedge\partial_{p_a}\quad(a,b=1,\dots,n),
\end{equation}
where $\{\partial_{q^a},\partial_{p_a}\}$ is the coordinate basis (for vector fields $X:U\subseteq\tilde\Gamma\to T\tilde\Gamma$) corresponding to the chart $(U,z)$ (the associated dual basis being denoted by $\{\mathrm{d}q^a,\mathrm{d}p_a\}$). The above relation implies that $\theta^{ab}$ should be the components of an antisymmetric tensor.

Now, the solution curves $t\mapsto c(t)$ to the equations of motion associated to the mechanical system $(\tilde\Gamma, \mathcal{H})$ are given, as we know, by the flow of the Hamiltonian vector field $X_\mathcal{H}$ (taking care of computing $X_\mathcal{H}$ with $\tilde{\mathsf{P}}$).  
The equations of motions are therefore \footnote{The mapping $T^\star\tilde\Gamma\ni\eta\mapsto\tilde{\mathsf{P}}(\eta,\cdot)=-\tilde{\mathsf{P}}(\cdot,\eta)\in T\tilde\Gamma$ is the inverse of the natural isomorphism $\tilde\omega^\flat:T\tilde\Gamma\to T^\star\tilde\Gamma$ (analogous to the mapping which lowers indices of tensor fields on a Riemannian manifold) provided by the symplectic structure $\tilde\omega$. It is sometimes denoted by $\tilde\omega^\sharp=(\tilde\omega^\flat)^{-1}$.}
\begin{equation}
\dot{c}=X_\mathcal{H}\circ c\qquad (X_{\mathcal{H}}=\tilde{\mathsf{P}}(\cdot,\mathrm{d}\mathcal{H})).\label{eqs-motion}
\end{equation}

We stress that, due to $(U,z)$ not being a symplectic chart, the above equations will not take the standard canonical form when written out explicitly in such patch. When working on a symplectic patch, say $(U',z')$, we can immediately make use of the standard form of Hamilton's equations but we should take care of writing the Hamiltonian $\mathcal{H}$ in such canonical coordinates $z'$. In this canonical representation, the main difference in systems $(\tilde\Gamma,\mathcal H)$ and $(\Gamma,\mathcal H)$ resides in the form taken by the Hamiltonian function (since both systems are written in patches in which the symplectic structure takes the flat standard form). Therefore, \emph{replacing} $\mathcal{H}_z$ by $\mathcal{H}_{z'}$ in the \emph{canonical} description of $(\Gamma,\mathcal{H})$ is tantamount to going over to the system $(\tilde\Gamma,\mathcal H)$.

The above discussion opens the possibility of interpreting $(\tilde\Gamma,\mathcal{H})$ as a ``deformation'' of a mechanical system $(\Gamma,\mathcal{H})$. 
The ``original'' system $(\Gamma,\mathcal{H})$ being the model initially put forward for investigating a determined classical physical phenomena, but nonetheless incomplete (in the interpretation of noncommutative quantum mechanics), the accurate theoretical model being $(\tilde\Gamma,\mathcal{H})$ instead. Such improved model would not entail a radical departure from the theoretical description provided by $(\Gamma,\mathcal{H})$, and could therefore be overlooked (due to the noncommutativity scale---controlled by $\theta^{ab}$---being difficult to perceive). 

Put another way, the mechanical system $(\tilde\Gamma,\mathcal{H})$ is the direct classical limit (in the sense of the Dirac correspondence) of the supposedly correct quantum mechanical system defined by algebra \eqref{nc-q-com-rel} and the Hamiltonian operator $\hat{\mathcal{H}}$. As such, it is $(\tilde\Gamma,\mathcal{H})$ which is the correct theoretical model for the description of the classical counterpart. Therefore, $\theta-$deformed quantum mechanics not only can give corrections at the quantum level, but also yields corrections within the purely \emph{classical} domain.\footnote{This will clearly be the case if algebra \eqref{nc-q-com-rel} is associated to the standard procedure of quantization (based on the Dirac correspondence) of a classical mechanical system (defined on a symplectic manifold).}

Such reasoning is of course appropriate to any extension of standard quantum mechanics based on generalizing algebra \eqref{nc-q-com-rel}, as long as the Lie-bracket structure is preserved when constructing the corresponding classical system $(\tilde\Gamma,\mathcal{H})$ (preservation of such structure is tantamount to a well defined symplectic manifold $\tilde\Gamma$ on which the mechanical system is to be modeled). In the case of the simple $\theta-$deformation this is achieved (as we saw above) by demanding that $\theta^{ab}$ be (the components of) an antisymetric tensor. A central part of the present letter is to apply the above rationale in order to interpret the effective scheme of loop quantum cosmology as the direct classical limit of a certain noncommutative quantum mechanics (albeit somewhat more general than the $\theta-$deformed one).

The remaining part of the manuscript is organized as follows. In section II we show how the effective scheme of LQC of the Flat FLRW model is to be interpreted as a purely classical limit of a certain quantum model. Section III is devoted to a similar analysis for the effective LQC of the Kantowski-Sachs and Bianchi type I models; while some alternate prescriptions for the effective flat FLRW are addressed in Section IV. Some recent generalizations are examined in Section V.
Finally, Section VI is devoted to discussion. 

\section{The case of the standard effective flat FLRW model}
As stated in Section I, the effective approximation of LQC can be understood in a naive way by replacing $b$ with $\sin(\lambda b)/\lambda$ in the Hamiltonian \eqref{ham1}---hence replacing Hamiltonian \eqref{ham1} by Hamiltonian \eqref{ham2}. That it is just a replacement---and not a map in minisuperspace---is evident from the fact that no attention is paid to how the symplectic structure should be expressed upon taking $b\to\sin(\lambda b)/\lambda$.  

From the discussion given above in \ref{ncqm}, it is conjectured that this so called ``holonomization'' of the Hamiltonian makes the effective scheme of LQC a \emph{de facto} classical limit of some noncommutative quantum model. We now endeavor in explicitly showing that this is indeed the case.

Let $(\Gamma,\mathcal{H})$ denote the standard FLRW classical system defined by \eqref{PB} and \eqref{ham1} (in the following, the Hamiltonian constraint $\mathcal{H}=0$ is always assumed as part of the definition of the mechanical systems under scrutiny). Consider also a family of symplectic manifolds $\Gamma^{\ell}$ defined by the family of (invertible) Poisson structures
\begin{equation}
\mathsf{P}^\ell_z=4\pi G\gamma\sqrt{1-(\ell b)^2}\partial_b\wedge\partial_v+\partial_\phi\wedge\partial_{p_\phi},\label{PS}
\end{equation}
with $z=(z^1,\dots,z^4)=(b,\phi,v,p_\phi)$ the same coordinates used in the description of $(\Gamma,\mathcal{H})$. (Clearly, with regard to the family $(\Gamma^\ell,\mathcal H)$, $b$ must be restricted to take values in $(-\ell^{-1},\ell^{-1})$.) We remark that $\mathcal{H}$ has the form
\begin{equation}
\mathcal{H}_z=-\frac{3}{8\pi G\gamma^{2}}b^2 v+\frac{p^{2}_{\phi}}{2v}\label{ham-z}
\end{equation}
in coordinates $z$ for \emph{all mechanical systems} $(\Gamma,\mathcal H)$, $(\Gamma^{\ell},\mathcal H)$. Notice that under these circumstances the one-parameter family $(\Gamma^\ell,\mathcal{H})$ of mechanical systems reduce to $(\Gamma,\mathcal{H})$ upon taking $\ell\to0$ (since in such limit the $\mathsf {P}^\ell$ have all been reduced to $\mathsf P$, $b$ can now range over the whole real line).

The classical algebra associated to \eqref{PS} is 
\begin{align}
\{b,v\}:=\mathsf{P}^\ell_z(\mathrm{d}b,\mathrm{d}v)=4\pi G\gamma\sqrt{1-(\ell b)^2},\label{c-algebra1-1}\\
\{\phi,p_\phi\}:=\mathsf{P}^\ell_z(\mathrm{d}\phi,\mathrm{d}p_\phi)=1\label{c-algebra1-2}
\end{align}
which clearly indicates that coordinates $z$ are \emph{not} canonical for the family $(\Gamma^\ell,\mathcal{H})$ (although they are indeed canonical for $(\Gamma,\mathcal H)$). 

We consider now the new set of coordinates $z'=(b',\phi',v',p_\phi')$, 
\begin{align}
\left(b,\phi,v,p_\phi\right)\mapsto\left(\frac{\arcsin\left(\ell b\right)}{\ell},\phi,v,p_\phi\right)=\left(b',\phi',v',p_\phi'\right).\label{can-trans}
\end{align}
(Given the range of $b$ spelled out above, $b'$ can be set to range over the usual branch $(-\pi/2\ell,\pi/2\ell)$ so as to make \eqref{can-trans} a well defined phase space coordinate transformation---i.e. a diffeomorphism.) Of course, the map \eqref{can-trans} can be seen as a trivial extension from the change of coordinates $\Phi:(b,\phi)\mapsto(b',\phi')$ on the base manifold. However, note that such extension \eqref{can-trans} is \emph{not} the ``canonical point transformation'' associated to $\Phi$ (see, e.g. \cite{corben})---in the terminology of Ref. \cite{marsden}, the change of local bundle charts \eqref{can-trans} is \emph{not} the ``lift'' $T^\star\Phi$ of the base manifold diffeomorphism $\Phi$---and so the representation of the symplectic structure could change when using coordinates $z'$.

In fact, in such patch $(U',z')$, the family of Poisson structures $\mathsf{P}^\ell$ takes the flat form
\begin{equation}
\mathsf{P}^\ell_{z'}=4\pi G\gamma\,\partial_{b'}\wedge\partial_{v'}+\partial_{\phi'}\wedge\partial_{p'_\phi}
\end{equation}
and so the classical algebra is now represented by the standard relations
\begin{align}
\{b',v'\}:=\mathsf{P}^\ell_{z'}(\mathrm{d}b',\mathrm{d}v')=4\pi G\gamma,\label{c-algebra2-1}\\
\{\phi',p_\phi'\}:=\mathsf{P}^\ell_{z'}(\mathrm{d}\phi',\mathrm{d}p'_{\phi})=1.\label{c-algebra2-2}
\end{align}
In other words, the patch $(U',z')$ is symplectic.\footnote{Strictly speaking, for the local trivialization to be symplectic it is mandatory that the constant factor $4\pi G\gamma$---featured in the purely gravitational part of the symplectic structure---be unity (which is not the case). As is customary in the LQC literature, we will allow ourselves to be a bit sloopy and still call symplectic to any chart of this type (and to any other related to the standard ``flat'' structures arising in connection with LQC).} 

On the other hand, the Hamiltonian is expressed in coordinates $z'$ as, 
\begin{equation}
\mathcal{H}_{z'}=-\frac{3}{8\pi G\ell^2\lambda^2}\sin^2(\ell b')v'+\frac{p_\phi'^2}{2v'}.\label{ham-z'}
\end{equation}

It suffices to pick the particular member $(\Gamma^{\lambda},\mathcal H):=(\Gamma^{\ell=\lambda},\mathcal H)$ of the family $(\Gamma^{\ell},\mathcal H)$ of dynamical systems to immediately arrive at the effective LQC scheme.

We then have accomplished the task of showing that the standard effective LQC scheme is just a---rather elementary---symplectic modification of the usual flat FLRW model $(\Gamma,\mathcal{H})$. 
%
%
%


\section{Some relevant anisotropic models}
\subsection{The case of the effective Kantwoski-Sachs model}
The classical Kantowski-Sachs model, when written in connectoin-triad variables, is defined by the dynamical system $(\mathcal{H}^{KS},\Gamma_{\mathrm{KS}})$, with the Hamiltonian given by (see, e.g. \cite{KS-ashtekar})
\begin{equation}
{\mathcal{H}^{\mathrm{KS}}}_z=-\frac{N}{2G\gamma^2}\left[2bc\sqrt{p_c}+(b^2+\gamma^2)\frac{p_b}{\sqrt{p_c}}\right]\label{class-KS}
\end{equation}
and with
\begin{equation}
{\mathsf{P}_{\mathrm{KS}}}_z= G\gamma\left(\partial_b\wedge\partial_{p_b}+2\partial_c\wedge\partial_{p_c}\right) 
\end{equation}
being the Poisson structure on $\Gamma_{\mathrm{KS}}$ (where ...). In the gauge $N=\gamma\sqrt p_c/b$ the above Hamiltonian is
\begin{equation}
{\mathcal{H}^{\mathrm{KS}}}_z=-\frac{1}{2G\gamma}\left[2cp_c+\left(b+\frac{\gamma^{2}}{b}\right)p_{b}\right].\label{class-KS2}
\end{equation}

The importance of this anisotropic model for quantum cosmology relies on the fact that it gives the description of the interior of the Schwarszchild black hole.

The orginal Corichi-Singh prescription \cite{KS-corichi}, as well as the improved Ashtekar-Olmedo-Singh one \cite{KS-ashtekar}, for the effective Kantowski-Sachs model is realized by making the replacement
\begin{equation}
b\to\frac{\sin\left(\lambda_1 b\right)}{\lambda_1},\quad c\to\frac{\sin\left(\lambda_2 c\right)}{\lambda_2}
\end{equation}
in Hamiltonian \eqref{class-KS}, where $\dot{\lambda_1}=\dot{\lambda_2}=0$
---for details see \cite{KS-corichi} and \cite{KS-ashtekar} (and references therein)---and with $b$ and $c$ taking values on $(0,\pi/\lambda_1)$ and $(0,\pi/\lambda_2)$ (alternatively, on $(-\pi/2\lambda_1,\pi/2\lambda_1)$ and $(-\pi/2\lambda_2,\pi/2\lambda_2)$ , respectively). The effective scheme is therefore defined by Hamiltonian 
\begin{equation}
{\mathcal{H}^{\mathrm{KS}}_{\mathrm{eff}}}_z=-\frac{1}{2G\gamma}\left[2\frac{\sin(\lambda_2 c)}{\lambda_2}p_c+\left(\frac{\sin(\lambda_1 b)}{\lambda_1}+\frac{\gamma^2\lambda_1}{\sin(\lambda_1 b)}\right)p_b\right]
\end{equation}
where the gauge $N=\gamma\delta\sqrt{p_c}/\sin(\delta b)$---which is the effective analog of $N=\gamma\sqrt p_c/b$---has been taken.

Consider now a family of dynamical systems $(\mathcal{H}^{\mathrm{KS}},\Gamma_{\mathrm{KS}}^{\ell_1,\ell_2})$, where the Poisson structure on $\Gamma_{\mathrm{KS}}^{\ell_1,\ell_2}$ is given by 
\begin{equation}
{\mathsf{P}^{\ell_1,\ell_2}_{\mathrm{KS}}}_{z}=G\gamma\left(\sqrt{1-\left(\ell_1 b\right)^2}\partial_{b}\wedge\partial_{p_b}+2\sqrt{1-\left(\ell_2 c\right)^2}\partial_{c}\wedge\partial_{p_c}\right),\label{PS_KS}
\end{equation}
where $b$ and $c$ must take values in $(-\ell_1^{-1},\ell_1^{-1})$ and $(-\ell_2^{-1},\ell_2^{-1})$, respectively, in order for $\mathsf{P}^{\ell_1,\ell_2}_{\mathrm{kS}}$ to comprise a two-parameter family of well defined symplectic manifolds.
Note that coordinates $z=(b,c,p_b,p_c)$ do \emph{not} define a symplectic patch for the family $\Gamma^{\ell_1,\ell_2}_{\mathrm{KS}}$. Indeed, the classical algebra associated to \eqref{PS_KS} is 
\begin{align}
\{b,p_b\}:=\mathsf{P}^{\ell_1,\ell_2}_z(\mathrm{d}b,\mathrm{d}p_b)= G\gamma\sqrt{1-(\ell_1 b)^2},\label{KS-algebra1-1}\\
\{c,p_c\}:=\mathsf{P}^{\ell_1,\ell_2}_z(\mathrm{d}c,\mathrm{d}p_c)=2 G\gamma\sqrt{1-(\ell_2 c)^2}.\label{KS-algebra1-2}
\end{align}

Consider now the change of local trivialization on $\Gamma^{\ell_1,\ell_2}_{\mathrm{KS}}$ defined by 
\begin{align}
\left(b,c,p_b,p_c\right)\mapsto\nonumber\\
\left(\frac{\arcsin\left(\ell_1 b\right)}{\ell_1},\frac{\arcsin\left(\ell_1 c\right)}{\ell_2},p_b,p_c\right)=\left(b',c',p_b',p_c'\right),\label{can-trans-KS}
\end{align}
where $b'$ and $c'$ can be set to take values in the usual branches $(-\pi/2\ell_1,\pi/2\ell_1)$ and $(-\pi/2\ell_2,\pi/2\ell_2)$, respectively (in view of the given ranges for $b$ and $c$). 

It can be easily shown that in coordinates $z'=(b',c',p_b',p_c')$ the Poisson structures $\mathsf{P}^{\ell_1,\ell_2}_{\mathrm{KS}}$ take always the corresponding flat form, i.e.
\begin{equation}
{\mathsf{P}_{\mathrm{KS}}^{\ell_1,\ell_2}}_{z'}= G\gamma\left(\partial_{b'}\wedge\partial_{p_b'}+2\partial_c'\wedge\partial_{p_c'}\right).
\end{equation}
In other words, coordinates $z'$ do define a symplectic patch for the family $\Gamma^{\ell}_{\mathrm{KS}}$. 

Now, the Hamiltonian $\mathcal{H}^{\mathrm{KS}}$ in variables $z'$ is written as
\begin{equation}
\mathcal{H}^{\mathrm{KS}}_{z'}=-\frac{1}{2G\gamma}\left[2\frac{\sin(\ell_2 c')}{\ell_2}p_c'+\left(\frac{\sin(\ell_1 b')}{\ell_1}+\frac{\gamma^2\ell_1}{\sin(\ell_1 b')}\right)p_b'\right]
\end{equation}
It is clear that the dynamical system $\left((\Gamma^{\lambda_1,\lambda_2}_{\mathrm{KS}},\mathcal{H}^{\mathrm{KS}}\right)$ (which is a particular element of the family $\left(\Gamma^{\ell_1,\ell_2}_{\mathrm{KS}},\mathcal{H}^{\mathrm{KS}}\right)$) is identical to the effective scheme of the Kantowski-Sachs model.

\subsection{The case of the effective Bianchi I model}\label{BI}
The classical Bianchi type I model, in the pressence of a free homogenous scalar field, is defined (in connection-triad variables) by the dynamical system $(\mathcal{H}^{\mathrm{BI}},\Gamma_{\mathrm{BI}})$, with the Hamiltonian given by
\begin{equation}
{\mathcal{H}^{\mathrm{BI}}}_z=-\frac{N}{8\pi G\gamma^2p_1p_2p_3}\left(c_1c_2p_1p_2+c_1c_3p_1p_3+c_2c_3p_2p_3\right)+\frac{p_\phi^2}{2}\label{class-BI}
\end{equation}
and 
\begin{equation}
{\mathsf{P}_{\mathrm{BI}}}_z=4\pi G\gamma\left(\partial_1\wedge\partial_{p_1}+\partial_2\wedge\partial_{p_2}+\partial_3\wedge\partial_{p_3}\right)+\partial_\phi\wedge\partial_{p_\phi} 
\end{equation}
the Poisson structure on $\Gamma_{\mathrm{BI}}$. 

The importance of this anisotropic model for quantum cosmology is based on the fact that its isotropic limit is the flat FLRW model. Additionally, it is prominently featured in the so-called BKL scenario \cite{BKL} (a recent monograph on the subject is \cite{BKL-book}).

We first tackle the somewhat simpler prescription akin to the polymer representation of the Bianchi I cosmology \cite{lqc-review2}. We therefore consider the replacement
\begin{equation}
c_1\to\frac{\sin(\lambda_1c_1)}{\lambda_1},\quad
c_2\to\frac{\sin(\lambda_2c_2)}{\lambda_2},\quad 
c_3\to\frac{\sin(\lambda_3c_3)}{\lambda_3}
\end{equation}
in Hamiltonian \eqref{class-BI}, where $\lambda_1$, $\lambda_2$, $\lambda_3$ are constants associated to the three independent polymer lattices (see \cite{lqc-review2}) and with $c_i$ taking values in $\left(0,\pi/\lambda_i\right)$  (alternatively, in $(-\pi/2\lambda_i,\pi/2\lambda_i$ )---for $i=1,2,3$. The effective Hamiltonian is therefore given by 
\begin{align}
{\mathcal{H}^{\mathrm{BI}}_{\mathrm{eff}}}_z=-\frac{1}{8\pi G\gamma^2}
\left(
\frac{\sin(\lambda_1c_1)}{\lambda_1}\frac{\sin(\lambda_2c_2)}{\lambda_2}p_1p_2\right.\nonumber\\
\left.+\frac{\sin(\lambda_1c_1)}{\lambda_1}\frac{\sin(\lambda_3c_3)}{\lambda_2}p_1p_3+
\frac{\sin(\lambda_2c_2)}{\lambda_2}\frac{\sin(\lambda_3c_3)}{\lambda_3}p_2p_3
\right)+\frac{p_\phi^2}{2}\label{eff-BI}
\end{align}
where we have taken the gauge $N=p_1p_2p_3$.

Consider now a family of dynamical systems $\left(\Gamma_{\mathrm{BI}}^{\ell_1,\ell_2,\ell_3},\mathcal{H}^{\mathrm{BI}}\right)$, where the Poisson structure on $\Gamma_{\mathrm{BI}}^{\ell_1,\ell_2,\ell_3}$ is given by 
\begin{align}
{\mathsf{P}^{\ell_1,\ell_2,\ell_3}_{\mathrm{BI}}}_{z}=&4\pi G\gamma\left(\sqrt{1-\left(\ell_1 c_1\right)^2}\partial_{1}\wedge\partial_{p_1}+\sqrt{1-\left(\ell_2 c_2\right)^2}\partial_{2}\wedge\partial_{p_2}\right.\nonumber\\
&\left.+\sqrt{1-\left(\ell_2 c\right)^2}\partial_{3}\wedge\partial_{p_3}\right)+\partial_\phi\wedge\partial_{p_\phi},\label{PS_BI}
\end{align}
where $c_i$ must take values in $(-\ell_i^{-1},\ell_i^{-1})$ ($i=1,2,3$).
Note that coordinates $z=(c_1,c_2,c_3,\phi,p_1,p_2,p_3,p_{\phi})$ \emph{do not} define a symplectic patch for the family $\Gamma^{\ell_1,\ell_2\ell_3}_{\mathrm{BI}}$. 

Consider now the change of local trivialization on $\Gamma^{\ell_1,\ell_2\ell_3}_{\mathrm{BI}}$ defined by 
\begin{align}
\left(c_1,c_2,c_3,\phi,p_1,p_2,p_3,p_{\phi}\right)\mapsto\nonumber\\
\left(\frac{\arcsin\left(\ell_1 c_1\right)}{\ell_1},\frac{\arcsin\left(\ell_2 c_2\right)}{\ell_2},\frac{\arcsin\left(\ell_3 c_3\right)}{\ell_3},\phi,p_1,p_2,p_3,p_{\phi}\right)\nonumber\\
=\left(c_1',c_2',c_3',\phi',p_1',p_2',p_3',p_{\phi}'\right).\label{can-trans-BI}
\end{align}
where $c_i'$ can be set to take values in the usual branch $(-\pi/2\ell_i,\pi/2\ell_i)$ (for $i=1,2,3$). 

It can be easily shown that in coordinates $z'=\left(c_1',c_2',c_3',\phi',p_1',p_2',p_3',p_{\phi}'\right)$ all the Poisson structures $\mathsf{P}^{\ell_1,\ell_2,\ell_3}_{\mathrm{BI}}$ take always the standard flat form, i.e.
\begin{equation}
{\mathsf{P}_{\mathrm{BI}}^{\ell_1,\ell_2,\ell_3}}_{z'}=4\pi G\gamma\left(\partial_{1'}\wedge\partial_{p_1'}+\partial_{2'}\wedge\partial_{p_2'}+\partial_{3'}\wedge\partial_{p_3'}\right)+\partial_\phi'\wedge\partial_{p'_{\phi}}.
\end{equation}
In other words, coordinates $z'$ do define a symplectic patch for the family $\Gamma^{\ell_1,\ell_2,\ell_3}_{\mathrm{BI}}$. 

Now, the Hamiltonian $\mathcal{H}^{\mathrm{BI}}$ in variables $z'$ (and in the gauge $N=p_1p_2p_3$) is written as
\begin{align}
\mathcal{H}^{\mathrm{BI}}_{z'}=-\frac{1}{8\pi G\gamma^2}
\left(
\frac{\sin(\ell_1c_1')}{\ell_1}\frac{\sin(\ell_2c_2')}{\ell_2}p_1'p_2'\right.\nonumber\\
\left.+\frac{\sin(\ell_1c_1')}{\ell_1}\frac{\sin(\ell_3c_3')}{\ell_2}p_1'p_3'+
\frac{\sin(\ell_2c_2')}{\ell_2}\frac{\sin(\ell_3c_3')}{\ell_3}p_2'p_3'
\right)+\frac{{p'_\phi}^2}{2}.\label{def-BI}
\end{align}
It is hence clear that the dynamical system $\left(\Gamma^{\lambda_1,\lambda_2,\lambda_3}_{\mathrm{BI}},\mathcal{H}^{\mathrm{BI}}\right)$---which is a particular system pertaining to the family $\left(\Gamma^{\ell_1,\ell_2,\ell_3}_{\mathrm{BI}},\mathcal{H}^{\mathrm{BI}}\right)$---is identical to the polymer Bianchi I cosmology.

We turn now to the improved $\bar{\mu}$ effective scheme (for details, see the reviews \cite{lqc-review, lqc-review2} and references therein), which is defined by the replacement
\begin{equation}
c_1\to\frac{\sin(\bar\mu_1c_1)}{\bar\mu_1},\quad
c_2\to\frac{\sin(\bar\mu_2c_2)}{\bar\mu_2},\quad c_3\to\frac{\sin(\bar\mu_3c_3)}{\bar\mu_3}
\end{equation}
within the classical Hamiltonian \eqref{class-BI}, where 
\begin{equation}
\bar{\mu}_1=\lambda\sqrt{\frac{p_1}{p_2p_3}},
\quad\bar{\mu}_2=\lambda\sqrt{\frac{p_2}{p_1p_3}},
\quad\bar{\mu}_3=\lambda\sqrt{\frac{p_3}{p_1p_2}},
\end{equation}
resulting in the improved-dynamics Hamiltonian ($N=p_1p_2p_3$)
\begin{align}
{\mathcal{H}^{\mathrm{BI}}_{\mathrm{imp}}}_z=-\frac{1}{8\pi G\gamma^2}
\left(
\frac{\sin(\bar\mu_1c_1)}{\bar\mu_1}\frac{\sin(\bar\mu_2c_2)}{\bar\mu_2}p_1p_2\right.\nonumber\\
\left.+\frac{\sin(\bar\mu_1c_1)}{\bar\mu_1}\frac{\sin(\bar\mu_3c_3)}{\bar\mu_2}p_1p_3+
\frac{\sin(\bar\mu_2c_2)}{\bar\mu_2}\frac{\sin(\bar\mu_3c_3)}{\bar\mu_3}p_2p_3
\right)+\frac{p_\phi^2}{2}.\label{eff2-BI}
\end{align}

Given the notable similarities with the previous cases, we attempt the same strategy. 

Consider therefore a family of dynamical systems $\left(\Gamma_{\mathrm{BI}}^{\,\mu},\mathcal{H}^{\mathrm{BI}}\right)$, where the Poisson structures on $\Gamma_{\mathrm{BI}}^{\,\mu}$ are given by 
\begin{align}
{\mathsf{P}^{\,\mu}_{\mathrm{BI}}}_{z}=&4\pi G\gamma\left(\sqrt{1-\left(\mu_1 c_1\right)^2}\partial_{1}\wedge\partial_{p_1}+\sqrt{1-\left(\mu_2 c_2\right)^2}\partial_{2}\wedge\partial_{p_2}\right.\nonumber\\
&\left.+\sqrt{1-\left(\mu_3 c_3\right)^2}\partial_{3}\wedge\partial_{p_3}\right)+\partial_\phi\wedge\partial_{p_\phi},\label{PS_BI}
\end{align}
where $\mu_i=\ell\bar\mu_i$ and each of the $c_i$ must take values in its  corresponding interval $(-\mu_i^{-1},\mu_i^{-1})$.
Note that coordinates $z=(c_1,c_2,c_3,p_1,p_2,p_3)$ \emph{do not} define a symplectic patch for the family $\Gamma^{\,\mu}_{\mathrm{BI}}$. 

Consider now the change of local trivialization on $\Gamma^{\,\mu}_{\mathrm{BI}}$ defined by 
\begin{align}
\left(c_1,c_2,c_3,\phi,p_1,p_2,p_3,p_\phi\right)\mapsto\nonumber\\
\left(\frac{\arcsin\left(\bar\mu_1 c_1\right)}{\bar\mu_1},\frac{\arcsin\left(\bar\mu_2 c_2\right)}{\bar\mu_2},\frac{\arcsin\left(\bar\mu_3 c_3\right)}{\bar\mu_3},\phi,p_1,p_2,p_3,p_\phi\right)\nonumber\\
=\left(c_1',c_2',c_3',\phi',p_1',p_2',p_3',p'_\phi\right).\label{can-trans-BI}
\end{align}
where $c_i'$ can be set to take values in the usual branch $(-\pi/2\bar\mu_i,\pi/2\bar\mu_i)$ (for $i=1,2,3$). 

Although in this case the prefactor of each $c_i$ now always depends on all the $p_i$ (in sharp contrast with the previous cases), by careful computation it can be shown that coordinates $z'=(c_1',c_2',c_3',\phi',p_1',p_2',p_3',p'_\phi)$ do comprise a symplectic patch for the family $\Gamma^{\,\mu}_{\mathrm{BI}}$.  That is, again all the $\mathsf{P}^{\,\mu}_{\mathrm{BI}}$ take always the standard flat form,
\begin{equation}
{\mathsf{P}_{\mathrm{BI}}^{\,\mu}}_{z'}=4\pi G\gamma\left(\partial_{1'}\wedge\partial_{p_1'}+\partial_{2'}\wedge\partial_{p_2'}+\partial_{3'}\wedge\partial_{p_3'}\right)+\partial_\phi'\wedge\partial_{p'_\phi}.
\end{equation}  
in such coordinates.

Now, the Hamiltonian $\mathcal{H}^{\mathrm{BI}}$ in variables $z'$ is written as
\begin{align}
\mathcal{H}^{\mathrm{BI}}_{z'}=-\frac{1}{8\pi G\gamma^2}
\left(
\frac{\sin(\ell_1c_1')}{\ell_1}\frac{\sin(\ell_2c_2')}{\ell_2}p_1'p_2'\right.\nonumber\\
\left.+\frac{\sin(\ell_1c_1')}{\ell_1}\frac{\sin(\ell_3c_3')}{\ell_2}p_1'p_3'+
\frac{\sin(\ell_2c_2')}{\ell_2}\frac{\sin(\ell_3c_3')}{\ell_3}p_2'p_3'
\right)+\frac{{p'_\phi}^2}{2}\label{def2-BI}
\end{align}
It is clear that by fixing $\ell=1$ (so that $\mu_i=\bar{\mu}_i$) the resulting particular member $\left(\Gamma^{\,\bar\mu}_{\mathrm{BI}},\mathcal{H}^{\mathrm{BI}}\right)$ is identical to the $\bar{\mu}$-scheme effective cosmology.
\section{Some alternative FLRW effective schemes}
\subsection{The mLQCI prescription}
The effective scheme of the first version of the modified LQC (mLQCI) is defined by \cite{yang}, \cite{dapor-liegener}, \cite{li} the Hamiltonian
\begin{equation}
{\mathcal{H}_{\mathrm{eff}}^{\mathrm{I}}}_z=-\frac{3v}{8\pi G\gamma^{2}}\left(\frac{\sin^{2}(\lambda b)}{\lambda^{2}}-(1+\gamma^2)\frac{\sin^{4}(\lambda b)}{\lambda^{2}}\right)+\frac{p^{2}_{\phi}}{2v}.
\end{equation}

This improved scheme has the advantage of...

We will endeavor in showing that this dynamical model can be also understood as a symplectic deformation of the standard FLRW classical system $\left(\Gamma,\mathcal{H}\right)$ (which is defined by \eqref{PB}, \eqref{ham1} and the Hamiltonian constraint).

Consider the family of dynamical systems $\left(\Gamma^m_{\mathrm{I}},\mathcal{H}\right)$ where the Hamiltonian (in coordinates $z=(b,v,\phi,p_\phi)$) is written as 
\begin{equation}
{\mathcal{H}}_{z}=-\frac{3v}{8\pi G\gamma^{2}}b^2+\frac{p^{2}_{\phi}}{2v},
\end{equation}
and where the Poisson structures are given by 
\begin{equation}
{\mathsf{P}^m_{\mathrm{I}}}_{z}=\left\{
\begin{matrix}
4\pi G\gamma\frac{b-2m^2(1+\gamma^2)b^3}{\sqrt{b^2-m^2\left(1+\gamma^2\right)b^4}}\partial_{b}\wedge\partial_{v}+\partial_{\phi}\wedge\partial_{p_{\phi}},&\, b\geq0\\
\phantom{aeg}&\phantom{aeg}\\
4\pi G\gamma\frac{b-2m^2(1+\gamma^2)b^3}{-\sqrt{b^2-m^2\left(1+\gamma^2\right)b^4}}\partial_{b}\wedge\partial_{v}+\partial_{\phi}\wedge\partial_{p_{\phi}},&\, b<0
\end{matrix}\right.,\label{PSI}
\end{equation}
with $b\in\left(-m^{-1}(1+\gamma^2)^{-1/2},m^{-1}(1+\gamma^2)^{-1/2}\right)$. Note that the family $\left(\Gamma^m_{\mathrm{I}},\mathcal{H}\right)$ reduces to the standard FLRW model $\left(\Gamma,\mathcal{H}\right)$ for $m\to0$.

The coordinate transformation
\begin{equation}(\bar{b},\bar{\phi},\bar{v},\bar{p}_\phi)=\left\{\begin{matrix}
\left(\sqrt{b^2-m^2(1+\gamma^2)b^4},\phi,v,p_\phi\right)&\bar{b}>0\\
\phantom{aeg}&\phantom{aeg}\\
\left(-\sqrt{b^2-m^2(1+\gamma^2)b^4},\phi,v,p_\phi\right)&\bar{b}<0
\end{matrix}
\right.
\end{equation}
allows to write the family $(\Gamma^{m}_{\mathrm{I}},\mathcal{H})$ in canonical form, with $\mathcal H$ being written as 
\begin{equation}
\mathcal{H}_{\bar{z}}=-\frac{3v}{8\pi G\gamma^{2}}\left[\bar{b}^2-m^2(1+\gamma^2)\bar{b}^4\right]+\frac{\bar{p}^{2}_{\phi}}{2\bar{v}}.
\end{equation}

Observe that $\mathcal{H}_{\bar{z}}$ is better suited for tackling the effective mLQCI model. Indeed, from the experience gained in the preceding sections, it is now easy to come up with the appropriate deformation which takes the family $\left(\Gamma^m,\mathcal{H}\right)$ to the effective mLQCI dynamical system: Replace the Poisson structures $\mathsf{P}^{m}_{\mathrm{I}}$ on the family $\Gamma^m_{\mathrm{I}}$ (which acquire the canonical form in coordinates $\bar{z}$) with
\begin{equation}
{\mathsf{P}^{\ell}_{\mathrm{I}}}_{\bar{z}}=4\pi G\gamma\sqrt{1-\left(\ell\bar{b}\right)^2}\partial_{\bar{b}}\wedge\partial_{\bar{v}}+\partial_{\bar{\phi}}\wedge\partial_{\bar{p}_{\phi}}\label{PSI-2}
\end{equation}
Let us denote by $(\Gamma^{\ell,m}_{\mathrm{I}},\mathcal{H})$ this new family of dynamical systems and refer to the subfamily with $m=\lambda$ as $\left(\Gamma^{\ell,\lambda}_{\mathrm{I}},\mathcal{H}\right)$. Also from the discussion on the previous sections, it is evident that the particular member $\left(\Gamma^{\lambda,\lambda}_{\mathrm{I}},\mathcal{H}\right)$ is equivalent to the mLQCI effective scheme. To see this, it suffices to use the symplectic patch defined by 
\begin{equation}
(b',\phi',v',p'_\phi)=\left(\frac{\arcsin(\ell\bar{b})}{\ell},\bar{\phi},\bar{v},\bar{p}_\phi\right)
\end{equation}
and then fix $\ell=\lambda$.

We have therefore managed to interpret the effective mLQCI scheme as two successive deformations of the standard FLRW classical model. It is not difficult to encode the two successive deformations into a single one. Let us write, for the sake of completeness, the general composite deformation :
\begin{align}
{\mathsf{P}^{\ell,m}_{\mathrm{I}}}_{z}=\nonumber\\
\left\{
\begin{matrix}
4\pi G\gamma\frac{\bar{b}-2m^2(1+\gamma^2)\bar{b}^3}{\sqrt{\bar{b}^2-m^2\left(1+\gamma^2\right)\bar{b}^4}}\sqrt{1-\left(\ell \bar{b}\right)^2}\partial_{b}\wedge\partial_{v}+\partial_{\phi}\wedge\partial_{p_{\phi}},&\, b\geq0\\
\phantom{aeg}&\phantom{aeg}\\
4\pi G\gamma\frac{\bar{b}-2m^2(1+\gamma^2)\bar{b}^3}{-\sqrt{\bar{b}^2-m^2\left(1+\gamma^2\right)\bar{b}^4}}\sqrt{1-\left(\ell \bar{b}\right)^2}\partial_{b}\wedge\partial_{v}+\partial_{\phi}\wedge\partial_{p_{\phi}},&\, b<0
\end{matrix}\right.,\label{PSI-3}
\end{align}
with $\bar{b}=\left(1+\sqrt{\frac{1-4m^2(1+\gamma^2)b^2}{2m^2(1+\gamma^2)}}\right)^{1/2}$.
%
%
\subsection{The mLQCII prescription}
The effective scheme of the second version of the modified LQC (mLQCII) is defined by \cite{yang}, \cite{dapor-liegener}, \cite{li} the Hamiltonian
\begin{equation}
{\mathcal{H}_{\mathrm{eff}}^{\mathrm{I}}}_z=-\frac{3v}{8\pi G\gamma^{2}}\left(\frac{\sin^{2}(\lambda b/2)}{\lambda^{2}}+\gamma^2\frac{\sin^{4}(\lambda b/2)}{\lambda^{2}}\right)+\frac{p^{2}_{\phi}}{2v}.
\end{equation}
Given the high similarity with the mLQCI scheme we can anticipate that the required deformations will be very similar to the ones required in addressing the mLQCI model. Indeed, consider firstly the family $(\Gamma^m_{\mathrm{II}},\mathcal{H})$ where the corresponding Poisson structures are 
\begin{equation}
{\mathsf{P}^m_{\mathrm{II}}}_z=\left\{
\begin{matrix}
4\pi G\gamma\frac{b+2m^2\gamma^2b^3}{\sqrt{b^2+m^2\gamma^2b^4}}\partial_{b}\wedge\partial_{v}+\partial_{\phi}\wedge\partial_{p_{\phi}},&\, b\geq0\\
\phantom{aeg}&\phantom{aeg}\\
4\pi G\gamma\frac{b+2m^2\gamma^2b^3}{-\sqrt{b^2+m^2\gamma^2b^4}}\partial_{b}\wedge\partial_{v}+\partial_{\phi}\wedge\partial_{p_{\phi}},&\, b<0
\end{matrix}\right..\label{PSII}
\end{equation}
When such family is written in the symplectic patch given by 
\begin{equation}(\bar{b},\bar{\phi},\bar{v},\bar{p}_\phi)=\left\{\begin{matrix}
\left(\sqrt{b^2+m^2\gamma^2b^4},\phi,v,p_\phi\right)&\bar{b}>0\\
\phantom{aeg}&\phantom{aeg}\\
\left(-\sqrt{b^2+m^2\gamma^2b^4},\phi,v,p_\phi\right)&\bar{b}<0
\end{matrix}
\right.
\end{equation}
the Hamiltonian takes the form 
\begin{equation}
\mathcal{H}_{\bar{z}}=-\frac{3v}{8\pi G\gamma^{2}}\left[\bar{b}^2+m^2\gamma^2\bar{b}^4\right]+\frac{\bar{p}^{2}_{\phi}}{2\bar{v}}.
\end{equation}
We can immediately observe that the additional required deformation amounts to replace the Poisson structures $\mathsf{P}^m_{\mathrm{II}}$ (which is in canonical form in coordinates $\bar{z}$) with
\begin{equation}
{\mathsf{P}^{\ell}_{\mathrm{II}}}_{\bar{z}}=4\pi G\gamma\sqrt{1-\left(\ell\bar{b}/2\right)^2}\partial_{\bar{b}}\wedge\partial_{\bar{v}}+\partial_{\bar{\phi}}\wedge\partial_{\bar{p}_{\phi}}.\label{PSII-2}
\end{equation}
Denoting by $(\Gamma^{\ell,m}_{\mathrm{II}},\mathcal{H})$ this extended family, it is now evident that the particular member $(\Gamma^{\lambda,\lambda}_{\mathrm{II}},\mathcal{H})$ is equivalent to the effective mLQCII scheme. In order to see this more transparently we only need to work in the symplectic patch defined by the coordinate transformation
\begin{equation}
(b',\phi',v',p'_\phi)=\left(\frac{\arcsin(\ell\bar{b})}{\ell},\bar{\phi},\bar{v},\bar{p}_\phi\right)
\end{equation}
and take $\ell=\lambda$. We have therefore succeded in interpreting the mLQCII effective scheme as a deformation of the standard FLRW classical model $(\Gamma,\mathcal{H})$.

As in the previous case, we write for completeness the full composite deformation:
\begin{align}
{\mathsf{P}^{\ell,m}_{\mathrm{II}}}_{z}=\nonumber\\
\left\{
\begin{matrix}
4\pi G\gamma\frac{\bar{b}+2m^2\gamma^2\bar{b}^3}{\sqrt{\bar{b}^2+m^2\gamma^2\bar{b}^4}}\sqrt{1-\left(\ell \bar{b}\right)^2}\partial_{b}\wedge\partial_{v}+\partial_{\phi}\wedge\partial_{p_{\phi}},&\, b\geq0\\
\phantom{aeg}&\phantom{aeg}\\
4\pi G\gamma\frac{\bar{b}+2m^2\gamma^2\bar{b}^3}{-\sqrt{\bar{b}^2+m^2\gamma^2\bar{b}^4}}\sqrt{1-\left(\ell \bar{b}\right)^2}\partial_{b}\wedge\partial_{v}+\partial_{\phi}\wedge\partial_{p_{\phi}},&\, b<0
\end{matrix}\right.,\label{PSII-3}
\end{align}
with $\bar{b}=\left(-1+\sqrt{\frac{1+4m^2\gamma^2b^2}{2m^2\gamma^2}}\right)^{1/2}$.
\section{Some recent generalizations}
We end the main part of this letter by considering some very recent generalizations to the effective LQC scheme. 

In Refs. \cite{nclqc1} and \cite{nclqc2} a noncommutative generalization (involving a momentum sector $\theta$-deformation) of the standard effective scheme of LQC (in the volume representation) was put forward and studied. In Ref. \cite{ncmlqc} such generalization was implemented in the modified LQC versions. It is worth mentioning that these more generic frameworks can be readily encompassed within the perspective given in the present letter. So that they, too, can be understood as mere symplectic modifications of the standard flat FLRW dynamical system.

For the standard effective scheme, it suffices to replace the Poisson structure $\mathsf{P}$ of the standard FLRW model $(\Gamma,\mathcal{H})$ with (recall that $z=(b,\phi,v,p_v)$)
\begin{equation}
{\mathsf{P}^\theta}_z=4\pi G\gamma\sqrt{1-(\ell b)^2}\partial_b\wedge\partial_v+\partial_\phi\wedge\partial_{p_\phi}+\theta\partial_{p_\phi}\wedge\partial_v,\label{ncelqcPS}
\end{equation}
in which, evidently, the first term accounts for the effective LQC contribution, whereas the second term encodes the said noncommutative generalization.

For the modified effective scheme I, it suffices to replace the Poisson structure $\mathsf{P}$ of the standard FLRW model $(\Gamma,\mathcal{H})$ with
\begin{align}
{\mathsf{P}_{\mathrm{I}}^\theta}_z=\nonumber\\
\left\{
\begin{matrix}
4\pi G\gamma\frac{\bar{b}-2m^2(1+\gamma^2)\bar{b}^3}{\sqrt{\bar{b}^2-m^2\left(1+\gamma^2\right)\bar{b}^4}}\sqrt{1-\left(\ell \bar{b}\right)^2}\partial_{b}\wedge\partial_{v}+\partial_{\phi}\wedge\partial_{p_{\phi}}+\theta\partial_{p_\phi}\wedge\partial_v,\\
\phantom{aeg}\\
4\pi G\gamma\frac{\bar{b}-2m^2(1+\gamma^2)\bar{b}^3}{-\sqrt{\bar{b}^2-m^2\left(1+\gamma^2\right)\bar{b}^4}}\sqrt{1-\left(\ell \bar{b}\right)^2}\partial_{b}\wedge\partial_{v}+\partial_{\phi}\wedge\partial_{p_{\phi}}+\theta\partial_{p_\phi}\wedge\partial_v
\end{matrix}\right.\label{ncmlqcPSI}
\end{align}
with $\bar{b}=\left(1+\sqrt{\frac{1-4m^2(1+\gamma^2)b^2}{2m^2(1+\gamma^2)}}\right)^{1/2}$.

For the modified effective scheme II, it suffices to replace the Poisson structure $\mathsf{P}$ of the standard FLRW model $(\Gamma,\mathcal{H})$ with
\begin{align}
{\mathsf{P}_{\mathrm{II}}^{\theta}}_z=\nonumber\\
\left\{
\begin{matrix}
4\pi G\gamma\frac{\bar{b}+2m^2\gamma^2\bar{b}^3}{\sqrt{\bar{b}^2+m^2\gamma^2\bar{b}^4}}\sqrt{1-\left(\ell \bar{b}\right)^2}\partial_{b}\wedge\partial_{v}+\partial_{\phi}\wedge\partial_{p_{\phi}}+\theta\partial_{p_\phi}\wedge\partial_v,\\
\phantom{aeg}\\
4\pi G\gamma\frac{\bar{b}+2m^2\gamma^2\bar{b}^3}{-\sqrt{\bar{b}^2+m^2\gamma^2\bar{b}^4}}\sqrt{1-\left(\ell \bar{b}\right)^2}\partial_{b}\wedge\partial_{v}+\partial_{\phi}\wedge\partial_{p_{\phi}}+\theta\partial_{p_\phi}\wedge\partial_v
\end{matrix}\right.\label{ncmlqcPSII}
\end{align}
with $\bar{b}=\left(-1+\sqrt{\frac{1+4m^2\gamma^2b^2}{2m^2\gamma^2}}\right)^{1/2}$. Again, for both noncommutative modified schemes, the first term accounts for the LQC contribution, while the second one gives the noncommutative extension.

As shown and discussed in Refs. \cite{nclqc1}, \cite{nclqc2}, \cite{ncmlqc}, the very simple additional momentum-sector term featured in the modified symplectic structures presented above are such that key features (e.g. the quantum bounce) of the LQC paradigm are conserved, with a pre-inflationary dynamics which in the overall matches very well that of effective LQC. At the same time, such simplistic modification gives rise to additional $\theta$-terms which might have a noticeable impact at a more subtle level (e.g. primordial perturbations), potentially allowing to alleviate some current tensions.

\section{Discussion}

It is very important to emphasize that although the pair $(\Gamma^\lambda,\mathcal{H})$ (together with the Hamiltonian constraint)---where the Hamiltonian \eqref{ham-z} is the one for the standard flat FLRW model and where the Poisson structure is defined by \eqref{PS}---is a dynamical system which is \emph{identical} to the one defined by the effective scheme of LQC, the former is merely a pure symplectic modification of the standard flat FLRW scalar cosmology system $(\Gamma,\mathcal H)$.

A similar situation was found to hold for the effective mLQCI and mLQCII models. Furthermore, the effective LQC of the anisotropic Kantowski-Sachs and Bianchi type I models were also shown to be pure symplectic modifications of their corresponding standard classical scenarios. Although both of the mLQC models were considered in their standard $\mu_0$-scheme form, the experience gained in tackling the $\bar{\mu}$-scheme of the Bianchi type I model (Section \ref{BI}) allows to confidently assert that the improved $\bar{\mu}$-scheme of both mLQC models (see \cite{li}) can also be understood in this way just by replacing the parameter $\ell$ in the Poisson structures \eqref{PSI-3} and \eqref{PSII-3} with the corresponding phase space function $\bar{\mu}$.
Taking advantage of this ``dictionary'' allows to depict different proposals for the effective approximation to LQC of highly relevant models as classical limits of some Wheeler-DeWitt quantum cosmology frameworks.

For instance, in view of the classical algebra \eqref{c-algebra1-1} and \eqref{c-algebra1-2}, the associated Wheeler-DeWitt framework is defined by the modified Heisenberg relations, 
\begin{equation}
[\hat{b},\,\hat{v}]=i4\pi G\gamma\hbar\widehat{\sqrt{1-(\lambda b)^2}},\quad[\hat{\phi},\,\hat{p}_\phi]=i\hbar\mathbf{1}.\label{q-algebra1}
\end{equation}
and the associated Wheeler-DeWitt equation 
\begin{equation}
\hat{\mathcal{H}}\psi(\hat{b},\hat{\phi})=0.
\end{equation}
Within the standard interpretation of the effective scheme of LQC, the question of seriously considering the above quantum paradigm is not to be regarded as a worthy endeavor, since such a scheme is already a kind of \emph{semiclassical} approximation to the full LQC \cite{lqc-review2}. However, as already stated, in the new vista portrayed above, the effective scheme of LQC is \emph{the exact classical limit} of a certain quantum theory. Therefore, within this interpretation, the quantization of such framework is in fact of paramount importance. Additionally, given the peculiarities underlying LQC, the resulting standard quantization of the purely classical model $(\Gamma^\lambda,\mathcal{H})$ ought to present meaningful---and perhaps interesting---departures from the full LQC framework.

\section*{Acknowledgments}
A. E. G. was partially supported by SNI-SECIHTI. 
		

\end{document}